\documentclass[superscriptaddress,twocolumn,amssymb,nofootinbib,10pt]{revtex4-2}
\usepackage[normalem]{ulem}
\usepackage[utf8]{inputenc}
\usepackage{comment}
\usepackage{graphicx}
\usepackage{graphics}
\usepackage{amsmath, amssymb}
\usepackage{graphics,bm}
\usepackage{graphicx}
\usepackage{bbold}
\usepackage{slashed}
\usepackage{feynmf}
\usepackage{wrapfig}
\usepackage{hyperref}
\usepackage{cancel}
\usepackage[usenames]{color}  

\usepackage[compat=1.1.0]{tikz-feynman}
\usepackage{tikz}
\usetikzlibrary{arrows,shapes}
\usetikzlibrary{trees}
\usetikzlibrary{matrix,arrows} 			
\usetikzlibrary{positioning}				
\usetikzlibrary{calc,through}				
\usepackage{pgffor}							

\newcommand{\re}[1]{{\color{red} #1}}

\newcommand{\be}{\begin{equation}}
\newcommand{\ee}{\end{equation}}

\newcommand{\bqa}{\begin{eqnarray}}
\newcommand{\eqa}{\end{eqnarray}}

\def\square{\vcenter{\vbox{\hrule height.4pt
          \hbox{\vrule width.4pt height4pt
          \kern4pt\vrule width.3pt}\hrule height.4pt}}}

\begin{document}

\title{Pion condensation in QCD at finite isospin density, the dilute Bose gas, and speedy Goldstone bosons}


\author{Jens O. Andersen}
\email{jens.andersen@ntnu.no}
\affiliation{Department of Physics, Faculty of Natural Sciences,NTNU, 
Norwegian University of Science and Technology, H{\o}gskoleringen 5,
N-7491 Trondheim, Norway}
\affiliation{Niels Bohr International Academy, Blegdamsvej 17, DK-2100 Copenhagen, Denmark}

\author{Qing Yu}
\email{yuq@swust.edu.cn}
\affiliation{School of Mathematics and Physics, Southwest University of Science and Technology, Mianyang 621010, China}

\author{Hua Zhou}
\email{zhouhua@cqu.edu.cn}
\affiliation{Department of Physics, Chongqing University, Chongqing 401331, People’s Republic of China}

\date{\today}

\begin{abstract}
We consider pion condensation in QCD at finite isospin density $\mu_I$ and zero temperature using two-flavor chiral perturbation theory ($\chi$PT).
The pressure is calculated to next-to-leading order (NLO) in the low-energy expansion. In the nonrelativistic limit,
we recover the classic result by Lee, Huang, and Yang for the energy density of a dilute Bose gas with an $s$-wave scattering 
length that includes loop corrections from $\chi$PT.
In the chiral limit, higher-order calculations are tractable.
We calculate the pressure to next-to-next-to-leading order (NNLO) in 
the low-energy expansion, which is an expansion in powers of $\mu_I^2/(4\pi)^2f^2$, where $f$ is the
(bare) pion decay constant.
The spontaneous breakdown of the global internal symmetry $U(1)_{I_3}$ gives rise to 
a massless Goldstone boson or phonon. We discuss the properties of the low-energy effective theory describing this mode.
Finally, we compare our results for the pressure and the speed of sound with recent
lattice simulations with 2+1 flavors. 
The agreement is very good for isospin chemical potentials up to 180-200 MeV, depending on the physical quantity.
\end{abstract}

\maketitle

\section{Introduction}
Bose-Einstein condensation (BEC), spontaneous symmetry breaking and the subsequent appearance of massless Goldstone modes frequently occur in condensed matter and
high-energy physics. In the past decades, there has been significant progress in our understanding
of the properties of these massless excitations, their classification, and the construction of low-energy effective theories that describe their dynamics, see e.g. Refs.~\cite{nielsen,kaon,tomas0,tomas,nicolis0,hidaka,nicolis12,nicolis1,nicolis2,nicolis3}.

The classic textbook example is the Bose-Einstein condensation 
of a dilute nonrelativistic Bose gas at zero and finite temperature~\cite{fetter}.
The homogeneous Bose gas has been studied extensively for many decades beginning with the paper by Bogoliubov~\cite{bogo} in the 1940s.
The nonrelativistic Lagrangian that describes the system at finite particle density is
\bqa
\nonumber
{\cal L}&=&\psi^{\dagger}(i\partial_0+\mu_{\rm NR})\psi-{1\over2m}\nabla\psi^{\dagger}\cdot\nabla\psi
-{1\over4}g(\psi^{\dagger}\psi)^2
\\ &&
-{1\over36}g_3(\psi^{\dagger}\psi)^3
+\cdots\;, 
\label{nrlag}
\eqa
where the quantum field $\psi^{\dagger}$ creates a particle, $\psi$ destroys a particle, $\mu_{\rm NR}$ is the nonrelativistic 
chemical potential, and $g$ and $g_3$ are coupling constants. 
The term $(\psi^{\dagger}\psi)^2$ represents two-particle scattering and the
coupling $g$ is related to the $s$-wave scattering length $a$ as $g={8\pi a\over m}$.
The term $(\psi^{\dagger}\psi)^3$ represents $3\rightarrow3$ scattering.
The dots indicate terms that are higher order in the fields $\psi$, $\psi^{\dagger}$
and/or  derivatives, and 
include all local terms that are consistent with the symmetries, for
example, Galilean invariance and the phase symmetry $\psi\rightarrow e^{i\phi}\psi$.
The latter continuous symmetry gives rise to the conservation of particle number.
The coefficients $g$, $g_3$, and the coefficients that multiply the higher-order terms can 
in principle be determined from the $n$-body potentials that describe interatomic interactions~\cite{braaten}. 

At zero temperature, the expansion parameter of the dilute Bose gas is the so-called 
(dimensionless)
gas parameter $\sqrt{na^3}$, where $n$ is the number density.
Bogoliubov~\cite{bogo} obtained the mean-field result for the energy density, 
${\cal E}(n)={2\pi n^2a\over m}$.
The leading correction to Bogoliubov's result for the energy density
was calculated by Lee, Huang, and Yang (LHY)~\cite{LN, LN0} for a hard-sphere potential.
Later, part of the next-to-leading order correction was calculated 
by Wu~\cite{wu}, by Hugenholz and Pines~\cite{hug}, and by Sawada~\cite{saw}. 
A complete next-to-leading result was obtained by Braaten and Nieto~\cite{braaten}
using effective-field theory methods.
The result depends not only on the scattering
length $a$ but also on an energy-independent term in the
scattering amplitude for $3\rightarrow3$ scattering. The result is
\bqa
{\cal E}(n)&=&{2\pi n^2a\over m}\left[1+{128\over15\sqrt{\pi}}\sqrt{na^3}
\right.\nonumber\\&&\left.
+\left({32\pi-24\sqrt{3}\over3}\log(na^3)+C\right)na^3\right]\;, 
\label{bosongas}
\eqa
where $C$ is a constant involving the coupling $g_3$.
The dependence of physical results on quantities other
than the $s$-wave scattering length
was already realized by Hugenholz and Pines. These effects are referred to as
nonuniversal effects and are mimicked by e.g. the 
term $g_3(\psi^{\dagger}\psi)^3$ in Eq.~(\ref{nrlag}).
Similarly, the effective range $r_s$ of the two-body potential can be included
by adding the operator ${1\over4}h[\nabla(\psi^{\dagger}\psi)]^2$.
The coupling $h$ is related to the effective range, $h=2\pi a^2r_s/m$. 
The leading correction to the results of LHY due to this term was calculated in Ref.~\cite{hermans} and is of higher order in the gas parameter than the terms shown in 
Eq.~(\ref{bosongas}).

Another example of Bose-Einstein condensate is pion condensation in QCD at finite isospin density.
Consider two-flavor QCD with two independent chemical potentials $\mu_u$ and $\mu_d$.
Instead of using the quark chemical potentials, we can express the Lagrangian
in terms of the baryon and isospin chemical potentials
$\mu_B={3\over2}(\mu_u+\mu_d)$ and $\mu_I=(\mu_u-\mu_d)$.
Normally, the phase diagram is shown in the $\mu_B$--$T$ plane, but by allowing for nonzero $\mu_I$, we can add a new axis, and the phases in the $\mu_I$--$T$ plane have received particular attention
due to the fact that QCD is free of the sign problem for $\mu_B=0$.
The system is therefore amenable to lattice simulations employing 
importance-sampling techniques. In a series of papers, Brandt et al. have mapped out the 
phase diagram in the $\mu_I$--$T$ plane, calculated the equation of state, the isospin density,
quark and pion condensates, and the speed of sound~\cite{gergy1,gergy2,gergy3,gergy4,pionstar,gergy5}.
The EoS has been used to model pion stars~\cite{pionstar}, which are compact objects consisting of
a pion condensate with leptons and neutrinos to ensure electric neutrality and weak equilibrium, first proposed in~\cite{carig2}.

Chiral perturbation theory ($\chi$PT)~\cite{wein,gasser1,gasser2} is a low-energy effective theory for the pseudo-Goldstone bosons that appear in QCD as a consequence of spontaneous symmetry breaking in the vacuum by the quark condensate. 
$\chi$PT at finite isospin chemical potential was first considered in Refs.~\cite{isostep,isostep2}.
The quark, pion, and axial condensates of QCD at finite isospin density and $T=0$
have been calculated at next-to-leading order in~\cite{jens1,jens2}.
The results are generally
in good agreement with lattice simulations with 2+1 flavors
for small values of the isospin chemical potential,
$\mu_I\ll 4\pi f$, where $\chi$PT is expected to be valid.
In this paper, we compare the pressure and the speed of sound including updated
lattice simulations with 2+1 flavor, that recently appeared in Ref.~\cite{gergy5}.

At first glance, the dilute Bose gas and Bose condensation in QCD may seem unrelated. However, it is known
that a relativistic Bose condensate for low densities, i.e. near the onset of Bose condensation~\footnote{The onset of BEC is exactly at $\mu_I=m_{\pi}$, where $m_{\pi}$ is the physical pion mass.}, behaves as a dilute Bose gas. It was explicitly shown in Ref.~\cite{sundby} that the pions in the pion stars mentioned above, to a good approximation are nonrelativistic, i.e. the pion density is very low.
In this paper, we show that at the next-to-leading order, the pion condensate behaves as a dilute Bose for low densities:
We reproduce the corrections to Bogoliubov's results
in the appropriate limit, namely for $\mu_I$ just above the physical pion mass.
In the final stages of this work, Ref.~\cite{newnic} appeared with some overlapping
results, for example recovering the LHY correction for interacting scalars.
We also consider the opposite limit, namely the ultrarelativistic or high-density regime.
In this regime, higher-order calculations are in fact tractable and we calculate the pressure to next-to-next-to-leading order in the low-energy expansion. Finally, we use Son's construction of the 
effective field theory for the phonons to obtain a low-energy description of the system.
The damping rate of the phonons scales as $p^5$ for small $p$, a result that was obtained by Beliaev long ago in the case of a weakly interacting Bose gas~\cite{beli}.

\section{Thermodynamics}
In the discussion below, the thermodynamic potential 
is a function of a chemical potential $\mu$ and a parameter that we for now denote by $\alpha$,
$\Omega=\Omega(\mu,\alpha)$.
In the case of a dilute Bose gas, the parameter $\alpha$ is identified with the 
order parameter $v$, which is the condensate density.
In chiral perturbation theory, the quark condensate is rotated into a pion condensate, specified
by the rotation angle $\alpha$, cf. Eq.~(\ref{ansatz}) below.
The Bose-condensed phase is characterized by a nonzero value of $\alpha$. 

The thermodynamic potential can be systematically expanded in powers of the gas parameter $\sqrt{na^3}$ in the dilute Bose gas or in powers of $p/f$ in a low-energy expansion in $\chi$PT  (see explanation in the beginning of Sec.~IV).
We can write this expansion as
\bqa
\Omega(\mu,\alpha)&=&\Omega_0(\mu,\alpha)+\Omega_1(\mu,\alpha)+\cdots\;, 
\eqa
where the subscript $n$ indicates the $n$-th order contribution in the expansion. 
The value of $\alpha$ that minimizes $\Omega(\mu,\alpha)$ 
is denoted by $\alpha^*$ and is found by solving 
\bqa
{\partial\Omega\over\partial\alpha}&=&0\;. 
\label{condition}
\eqa
The pressure ${\cal P}$ is given by minus the thermodynamic potential evaluated at its minimum as a function of $\alpha$
\bqa
\mathcal{P}(\mu)&=&-\Omega(\mu,\alpha^*)\;.
\eqa
The charge density $n_Q$ associated with $\mu$ is given by
\bqa
n_Q(\mu)&=&-{\partial\Omega\over\partial\mu}\bigg|_{\alpha=\alpha^*}
={d\mathcal{P}\over d\mu}
\;, 
\label{densitet}
\eqa
where we in the last step
have used Eq.~(\ref{condition}).
Finally, the energy density is given by a Legendre transform of the pressure,
\bqa
\label{legendre}
{\cal E}(n_{Q})&=&-\mathcal{P}(\mu)+\mu n_{Q}(\mu)\;.
\eqa
The solution $\alpha^*$ to Eq.~(\ref{condition}) can also be written as a series 
\bqa
\alpha^*&=&\alpha_0+\alpha_1+\cdots\;.
\eqa
We can find expressions for $\alpha_0$, $\alpha_1$...by 
expanding Eq.~(\ref{condition}) around $\alpha=\alpha_0$. We find
\bqa
\nonumber
{\partial\Omega\over\partial\alpha}\bigg|_{\alpha=\alpha^*}&=&{\partial\Omega_0\over\partial\alpha}\bigg|_{\alpha=\alpha_0}+
{\partial^2\Omega_0\over\partial\alpha^2}\bigg|_{\alpha=\alpha_0}\alpha_1+
{\partial\Omega_1\over\partial\alpha}\bigg|_{\alpha=\alpha_0}\\ &&
+\cdots=0\;. 
\label{expansion}
\eqa
$\alpha_0$ is simply given by the solution to ${\partial\Omega_0\over\partial\alpha}=0$, so the
first term in Eq.~(\ref{expansion}) vanishes. We can solve then solve Eq.~(\ref{expansion})
for $\alpha_1$,
\bqa
\alpha_1&=&-{\partial\Omega_1\over\partial\alpha}\bigg|_{\alpha=\alpha_0}\bigg/{\partial^2\Omega_0\over\partial\alpha^2}\bigg|_{\alpha=\alpha_0}\;. 
\eqa
Finally, the expansion of the pressure reads
\bqa
\nonumber
\mathcal{P}(\mu)&=&-\Omega(\mu,\alpha^*)
\\ \nonumber
&=&-\Omega_0(\mu,\alpha_0)
-{\partial\Omega_0\over\partial\alpha}\bigg|_{\alpha=\alpha_0}\alpha_1
-\Omega_1(\mu,\alpha_0)+\cdots
\\ 
&=&
-\Omega_0(\mu,\alpha_0)-\Omega_1(\mu,\alpha_0)+\cdots
\;.
\label{pressurerekkje}
\eqa

\section{Dilute Bose gas}
\label{dilute}
We now derive the first two terms in the expansion Eq.~(\ref{bosongas}).
The first term is the mean-field result, while the second arises from a one-loop calculation.
The complex field is written as $\psi=v+\tilde{\psi}$, where
$v=\langle\psi\rangle$ is its expectation value and $\tilde{\psi}$ is a fluctuating 
quantum field. The fluctuating field is written as
$\tilde{\psi}={1\over\sqrt{2}}(\psi_1+i\psi_2)$.
To second order in the fluctuations, one finds the different terms of the Lagrangian Eq.~({\ref{nrlag}}) 
\bqa
{\cal L}^{\rm static}
&=&
\mu_{\rm NR}v^2-{1\over4}gv^4\;, 
\\
{\cal L}^{\rm linear}&=&{vX\over\sqrt{2}m}\psi_1\;, \\
{\cal L}^{\rm quadratic}&=&{1\over2}(\dot{\psi}_1\psi_2-\psi_1\dot{\psi}_2)
+{1\over4m}\psi_1\left(\nabla^2+{Y}\right)\psi_1
\nonumber\\ &&
+{1\over4m}\psi_2\left(\nabla^2+{X}\right)\psi_2\;, 
\eqa
where $X=2m(\mu_{\rm NR}-{1\over2}gv^2)$ and $Y=2m(\mu_{\rm NR}-{3\over2}gv^2)$.
The propagator matrix is given by the inverse of the quadratic terms in 
${\cal L}$.
In momentum space, one finds
\bqa
D(P)
={i\over p_0^2-E^2(p)+i\epsilon}
\begin{pmatrix}
{1\over2m}\left(p^{2}-X\right)&-ip_{0}\\
ip_{0}&{1\over2m}\left(p^{2}-Y\right)\\
\end{pmatrix}\;, 
\eqa
where $P$ is the four momentum, $P=(p_0,{\bf p})$, $p=|{\bf p}|$, and 
the spectrum is
\bqa
\label{spectri}
E(p)&=&{1\over2m}\sqrt{(p^2-X)(p^2-Y)}\;. 
\eqa
The thermodynamic potential in the mean-field approximation is as usual given by minus the
static part of the Lagrangian,
\bqa
\Omega_0(\mu_{\rm NR},v)&=&-\mu_{\rm NR}v^2+{1\over4}gv^4\;. 
\eqa
The linear term vanishes at the minimum $v_0=\sqrt{{2\mu_{\rm NR}\over g}}$
of the thermodynamic potential $\Omega_0(\mu_{\rm NR},v)$.
At the minimum, Eq.~(\ref{spectri}) reduces to the Bogoliubov spectrum
$E_p={p\over2m}\sqrt{p^2+4m\mu_{\rm NR}}$.
The dispersion relation is gapless and linear for small momenta 
$p^2\ll 4m\mu_{\rm NR}$ and that of a free nonrelativistic particle for large momenta $p^2\gg 4m\mu_{\rm NR}$. The length scale $1/\sqrt{4m\mu_{\rm NR}}$ is referred to as the coherence length.
This is the Goldstone mode, which is the result of the spontaneous breaking of the $U(1)$ phase symmetry mentioned above.
The NLO pressure is given by the thermodynamic potential evaluated at the classical minimum
$v_0$ cf. Eq.~(\ref{pressurerekkje}). This is convenient since $X=0$.
The NLO pressure is  
\bqa
\nonumber
\mathcal{P}(\mu_{\rm NR})&=&-\Omega_{0}(\mu_{\rm NR},v_0)-\Omega_1(\mu_{\rm NR},v_0)
\\ \nonumber
&=&{\mu_{\rm NR}^2\over g}-{1\over2}\int_pE(p)
={\mu_{\rm NR}^2\over g}-
{1\over4m}I_{0,-1}(M^2)\\
&=&
{\mu_{\rm NR}^2\over g}\left[1-{16(4m)^{{3\over2}}\sqrt{\mu_{\rm NR}g^2}\over15(4\pi)^2}\right]\;. 
\eqa
where the integrals $I_{m,n}(M^2)$ are defined in Eq.~(\ref{defigam}), 
$I_{0,-1}(M^2)$ is defined in Eq.~(\ref{i01}),
and $M^2=4m\mu_{\rm NR}$. The density is then given by 
\bqa
\nonumber
n(\mu_{\rm NR})&=&
{d{\cal P}\over d\mu}=
{2\mu_{\rm NR}\over g}-{1\over2}I_{1,1}(M^2)\\
&=&{2\mu_{\rm NR}\over g}\left[1-{4(4m)^{3\over2}\sqrt{\mu_{\rm NR}g^2}\over3(4\pi)^2}\right]\;, 
\label{nnr}
\eqa
where we have used Eq.~(\ref{i11}) in the last line. We can invert Eq.~(\ref{nnr}) to obtain the chemical potential in terms of the
number density. To the order we are calculating, we obtain 
\bqa
\nonumber
\mu_{\rm NR}(n)&=&{1\over2}gn+{1\over4}gI_{1,1}(2mgn) \\
&=&
{4\pi na\over m}\left[1+{32\over3\sqrt{\pi}}\sqrt{na^3}\right]
\;. 
\eqa
The energy density is then
\bqa
\nonumber
{\cal E}(n)&=&-\mathcal{P}(\mu_{\rm NR})+\mu_{\rm NR}(n)n\;,\\
&=&{1\over4}gn^2+{1\over4m}I_{0,-1}(2mgn)\;, 
\eqa
where we have used the LO relation $2\mu_{\rm NR}=gn$ in the integral, which is correct to
the order we are calculating. Note that the terms involving $I_{1,1}(\mu_{\rm NR})$ 
cancel in the final result for the energy density. 
We obtain the result
of Lee, Huang, and Yang~\cite{LN,LN0},
\bqa
{\cal E}(n)&=&{2\pi n^2a\over m}\left[1+{128\over15\sqrt{\pi}}\sqrt{na^3}\right]\;.
\label{bosongas2}
\eqa
It is amusing to note that all the divergent integrals appearing in this calculation are finite
in dimensional regularization in the limit $d\rightarrow3$. 
The reason is simply that the UV divergences are power divergences which
are always set to zero. Using a more conventional three-dimensional cutoff $\Lambda$ requires
the renormalization of $g$ and $\mu$ to obtain the finite result above~\cite{braaten}.

The quantum field theory in Eq.~(\ref{nrlag}) is nonrenormalizable.
This implies that the UV divergences that show up in the calculations of e.g. the energy density
at higher loop orders cannot be removed by renormalizing the couplings 
of the terms involved in the calculation.
This is exactly what happens at NNLO in the
low-density expansion. Some of the relevant two-loop vacuum graphs arising from 
the operator $(\psi^{\dagger}\psi)^2$
have logarithmic divergences
proportional to $a^4n^3$. These divergences are canceled by the counterterm for $g_3$
that multiplies the operator $(\psi^{\dagger}\psi)^3$~\cite{braaten}.

\section{Chiral perturbation theory}
Chiral perturbation theory is a low-energy effective theory for QCD that describes the 
pseudo-Goldstone bosons \cite{wein,gasser1,gasser2}, where the  $SU(2)_L\times (SU(2)_R$ global symmetry of QCD 
(for two flavors) is realized nonlinearly.
The vacuum manifold is $SU(2)_L\times SU(2)_R/SU(2)_V\simeq SU(2)$ and 
parametrized as $\Sigma=e^{i\phi_a\tau_a/f}$, where $\phi_a$ are the Goldstone fields, $f$ is the
bare pion-decay constant, and $\tau_a$ are the broken generators.

The chiral Lagrangian has a systematic low-energy expansion. Each 
covariant derivative counts as one power of momentum $p$ and each quark mass term counts as two powers of $p$. 
Using this power-counting scheme, one writes down
all possible terms at each order in the expansion. At leading order, there
are two terms in the chiral Lagrangian. For two flavors, it reads
\bqa
{\cal L}_{2}&=&{1\over4}f^2\langle\nabla^{\mu}\Sigma^{\dagger}\nabla_{\mu}\Sigma\rangle+{1\over4}f^2\langle\chi^{\dagger}\Sigma+\Sigma^{\dagger}\chi\rangle\;,
\label{l2}
\eqa
where $\langle A\rangle$ denotes the trace of a $2\times2$ matrix $A$ and the covariant derivatives are
\bqa
\nabla_{\mu}\Sigma&=&\partial_{\mu}\Sigma-i\left[\upsilon_{\mu},\Sigma\right]\;,\\
\nabla_{\mu}\Sigma^{\dagger}&=&\partial_{\mu}\Sigma^{\dagger}-i\left[\upsilon_{\mu},\Sigma^{\dagger}\right]\;,
\eqa
where $\upsilon_{\mu}={1\over2}\mu_I\tau_3\delta_{\mu 0}$, $\mu_{I}$ is the isospin chemical potential, and $\chi=2B_0{\rm diag}(m_u,m_d)$. 
The constant $B_0$ is related to the tree-level values of the 
light quark condensates in the vacuum via
$\langle\bar{d}d\rangle=\langle\bar{u}u\rangle=-f^2B_0$.
In the remainder of this paper, we work in the
isospin limit, $m_u=m_d$. 

The ground state for two-flavor QCD 
is of the form $\Sigma=e^{i\phi_a\tau_a/f}$ for constant fields $\phi_a$.
It is convenient to reparametrize the constant fields defining $\hat{\phi}_a$ and $\alpha$
via $\phi_a=\hat{\phi}_a\alpha f$ with the constraint 
$\hat{\phi}_1^2+\hat{\phi}_2^2+\hat{\phi}_3^2=1$.
The ground state is now denoted by $\Sigma_{\alpha}=e^{i\hat{\phi}_a\tau_a\alpha}$
and is properly normalized, $\Sigma^{\dagger}_{\alpha}\Sigma_{\alpha}=\mathbb{1}$. 
In the QCD vacuum, the ground state $\Sigma_0$ corresponds to $\phi_a=\alpha=0$, which
is simply the unit matrix, $\Sigma_0=\mathbb{1}$.
Using the normalization of $\hat{\phi_a}$ and the properties of the Pauli matrices, 
we can write the ground state as
\bqa
\Sigma_{\alpha}&=&e^{i\hat{\phi}_a\tau_a\alpha}
=\mathbb{1}\cos\alpha+i\hat{\phi}_a\tau_a\sin\alpha\;.
\label{ansatz}
\eqa
The parameter $\alpha$ can be thought of as 
a rotation angle, where the QCD vacuum ($\alpha=0$)
with a quark condensate is rotated into a state with a nonzero pion condensate. 
We can further restrict the values of $\hat{\phi}_a$ by considering the 
static Hamiltonian ${\cal H}_{\rm static}$ corresponding to the static part of the Lagrangian 
Eq.~(\ref{l2}). It reads
\bqa
\nonumber
{\cal H}_{\rm static}&=&-{1\over4}f^2\langle\chi^{\dagger}\Sigma+\Sigma^{\dagger}\chi\rangle
+{1\over16}f^2\mu_I^2\langle[\Sigma^{\dagger},\tau_3][\Sigma,\tau_3]\rangle\;. 
\\ &&
\label{ham}
\eqa
Substituting Eq.~(\ref{ansatz}) into Eq.~(\ref{ham}), we obtain the ground-state energy density 
\bqa
\nonumber
\langle{\cal H}_{\rm static}\rangle
&=&-f^2m_{\pi,0}^2\cos\alpha-{1\over2}f^2\mu_I^2(\hat{\phi}_1^2+\hat{\phi}_2^2)\sin^2\alpha
\;. 
\\ &&
\eqa
Here $m_{\pi,0}$ is the tree-level pion mass, which satisfies $m^2_{\pi,0}=B_0(m_u+m_d)$. The first term is minimized for
$\alpha=0$. i.e. the QCD vacuum, while the second term is minimized for
$\alpha={1\over2}\pi$. 
The second term is minimized for $\hat{\phi}_1^2+\hat{\phi}_2^2=1$ and therefore $\hat{\phi}_3=0$. 
Without loss of generality
we can choose $\hat{\phi}_2=1$, which we do henceforth. The ground state can now be compactly written as
\bqa
\Sigma_{\alpha}&=&A_{\alpha}\Sigma_0A_{\alpha}\;, 
\eqa
with $A_{\alpha}=e^{{1\over2}i\tau_2\alpha}$.

We have rotated the ground state according to Eq.~(\ref{ansatz}), however, it turns out that the naive expression for $\Sigma$, $\Sigma=U\Sigma_{\alpha}U$,
where $U=e^{i{\phi_a\tau_a/2f}}$, is no longer valid~\cite{splitt}.
Instead of using $\Sigma=U\Sigma_{\alpha}U$,  we must use
\bqa
\Sigma&=&L_{\alpha}\Sigma_{\alpha}R_{\alpha}^{\dagger}\;,
\label{correct}
\eqa
with
\bqa
L_{\alpha}=A_{\alpha}UA_{\alpha}^{\dagger}\;,\quad
R_{\alpha}=A_{\alpha}^{\dagger}U^{\dagger}A_{\alpha}\;.
\eqa
The correct expression for $\Sigma$ then reads
\bqa
\label{correct2}
\Sigma&=&A_{\alpha}U^2A_{\alpha}
\;.
\eqa
\section{Thermodynamics to ${\cal O}(p^4)$}

Using the parametrization Eq.~(\ref{correct2}), we can expand the LO chiral Lagrangian
in powers of the fields $\phi_a$,
\sout{. To quartic order, we find}
\bqa
{\cal L}_{2}&=&{\cal L}^{\rm (0)}_{2}+{\cal L}^{\rm (1)}_{2}+{\cal L}^{\rm (2)}_{2}+{\cal L}^{\rm (3)}_{2}
+{\cal L}^{\rm (4)}_{2}+\cdots\;,
\eqa
where the superscript indicates the number of fields and where
\bqa
{\cal L}^{\rm (0)}_{2}&=&f^{2}m_{\pi,0}^{2}\cos\alpha+\frac{1}{2}f^{2}\mu^{2}_{I}\sin^{2}\alpha
\label{l0stat}
\;,\\
{\cal L}^{\rm (1)}_{2}&=&-fm_{\pi,0}^{2}\sin\alpha\phi_{2}+{f\mu^{2}_{I}\sin\alpha\cos\alpha}\phi_{2}\nonumber\\&&
-f\mu_{I}\sin\alpha\partial_{0}\phi_{1}\;,\\
{\cal L}^{\rm (2)}_{2}&=&\frac{1}{2}\partial^{\mu}\phi_{a}\partial_{\mu}\phi_{a}
+{1\over2}m_{12}(\phi_{1}\partial_{0}\phi_{2}-\phi_{2}\partial_{0}\phi_{1})\nonumber\\&&
-\frac{1}{2}m_a^2\phi_a^2\;,
\label{l2p2}\\
{\cal L}^{\rm (3)}_{2}&=&\frac{m_{\pi,0}^{2}\sin\alpha-4\mu^{2}_{I}\sin\alpha\cos\alpha}{6f}\phi_{2}\phi_{a}\phi_{a}\nonumber\\&&
+
\frac{\mu_{I}\sin\alpha}{f}\partial_{0}\phi_{1}(\phi^{2}_{2}+\phi^{2}_{3})\;,\\
{\cal L}^{\rm (4)}_{2}
&=&\frac{1}{24f^{2}}\phi_{a}\phi_{a}\left[(m_{\pi,0}^{2}\cos\alpha-4\mu^{2}_{I}\cos^{2}\alpha)\phi^{2}_{1}
\right.\nonumber\\&&\left.
+(m_{\pi,0}^{2}\cos\alpha-4\mu^{2}_{I}\cos2\alpha)\phi^{2}_{2}\right.
\nonumber\\&&\left.
+(m_{\pi,0}^{2}\cos\alpha+4\mu^{2}_{I}\sin^{2}\alpha)\phi^{2}_{3}\right]\nonumber\\&&
-\frac{\mu_{I}\cos\alpha}{3f^{2}}\phi_{a}\phi_{a}(\phi_{1}\partial_{0}\phi_{2}
-\phi_{2}\partial_{0}\phi_{1})\nonumber\\&&
+\frac{1}{6f^{2}}\left[\phi_{a}\phi_{b}\partial^{\mu}\phi_{a}\partial_{\mu}\phi_{b}
-\phi_{a}\phi_{a}\partial^{\mu}\phi_{b}\partial_{\mu}\phi_{b}\right]\;,
\label{l2p4}
\eqa
where the masses are
\bqa
m^{2}_{1}&=&m_{\pi,0}^{2}\cos\alpha-\mu^{2}_{I}\cos^{2}\alpha\;,\\
m^{2}_{2}&=&m_{\pi,0}^{2}\cos\alpha-\mu^{2}_{I}\cos2\alpha\;,\\
m^{2}_{3}&=&m_{\pi,0}^{2}\cos\alpha+\mu^{2}_{I}\sin^{2}\alpha\;,\\
m_{12}&=&2\mu_{I}\cos\alpha\;.
\eqa
From the quadratic term, the inverse propagator is
\bqa
D^{-1}(P)&=
\begin{pmatrix}
D^{-1}_{12}(P)&0\\
0&P^{2}-m_{3}^{2}
\end{pmatrix}\;,\\
D^{-1}_{12}(P)&=
\begin{pmatrix}
P^{2}-m_1^{2}&ip_{0}m_{12}\\
-ip_{0}m_{12}&P^{2}-m_2^{2}\\
\end{pmatrix}\;.
\eqa
The dispersion relations are found by solving $\det D^{-1}(P)=0$. One finds
\bqa
\nonumber
E_{\pm}^{2}(p)&=&p^2+{1\over2}(m_1^2+m_2^2+m_{12}^2)\\
&&\pm{1\over2}\sqrt{4p^2m_{12}^2+(m_1^2+m_2^2+m_{12}^2)^2-4m_1^2m_2^2}\;,
\nonumber\\&&\\
E_3^2(p)&=&p^2+m_3^2\;.
\eqa
At LO, the thermodynamic potential is minus the
static Lagrangian Eq.~(\ref{l0stat}),
\bqa
\label{loomega}
\Omega_0(\mu_I,\alpha)&=&-f^2m_{\pi,0}^2\cos\alpha-{1\over2}f^2\mu_I^2\sin^2\alpha\;.
\eqa
The value $\alpha_0$ that extremizes $\Omega_0(\mu_I,\alpha)$ satisfies 
$\cos\alpha_0=m_{\pi,0}^2/\mu_I^2$.
Note that the transition from the vacuum phase ($\alpha=0$) to the pion-condensed phase ($\alpha>0$)
takes place at $\mu_I=m_{\pi,0}$, i.e. the tree-level value of the physical pion mass $m_{\pi}$. The onset of BEC at the physical pion mass is expected to hold to all orders
in the low-energy expansion.
It follows from Eq.~(\ref{loomega}) that 
the pressure and the other thermodynamic
quantities are independent of $\mu_I$ for $\mu_I\in[0,m_{\pi,0}]$, i. e. before the onset of Bose condensation. This is an example of the 
Silver-Blaze property~\cite{cohen}.
Subtracting the constant pressure in the vacuum, the LO pressure in the BEC phase, 
\bqa
\mathcal{P}_0
&=&
{1\over2}f^2\mu_I^2\left[1-{m_{\pi,0}^2\over\mu_I^2}\right]^2\;,
\hspace{1cm}
\mu_I\geq m_{\pi,0}
\;.
\label{lop}
\eqa
At ${\cal O}(p^4)$, the chiral Lagrangian at finite chemical potential is~\cite{gasser1}
\bqa
\nonumber
{\cal L}_4&=&  {1\over4}l_1\langle\nabla_{\mu}\Sigma^{\dagger}\nabla^{\mu}\Sigma\rangle^2
+{1\over4}l_2\langle\nabla_{\mu}\Sigma^{\dagger}\nabla_{\nu}\Sigma\rangle
    \langle \nabla^{\mu}\Sigma^{\dagger}\nabla^{\nu}\Sigma\rangle
\\ && \nonumber
    +{1\over16}(l_3+l_4)\langle\chi^{\dagger}\Sigma+\Sigma^{\dagger}\chi\rangle^2
\\&&
    +{1\over8}{l_4}\langle\nabla_{\mu}\Sigma^{\dagger}\nabla^{\mu}\Sigma\rangle
    \langle\chi^{\dagger}\Sigma+\Sigma^{\dagger}\chi\rangle
+{1\over2}h_1\langle\chi^{\dagger}\chi\rangle\;,
\label{l4}
\eqa
where $l_1$--$l_4$ and $h_1$ are bare coupling constants.
In a next-to-leading order calculation, there are two contributions to the thermodynamic potential.
The first one is the one-loop functional determinant of bosonic fluctuations and the second is the static part of ${\cal L}_4$.
The functional determinant is regularized using dimensional regularization and contains poles
in $\epsilon$, where $d=3-2\epsilon$. These ultraviolet divergences are removed
by renormalizing the constants $l_i$ and $h_1$.
The relations between the bare couplings and their renormalized counterparts are
\bqa
\nonumber
l_i=l_i^r-{\gamma_i\Lambda^{-2\epsilon}\over2(4\pi)^2}\left[{1\over\epsilon}+1\right]\;,
h_i=h_i^r-{\delta_i\Lambda^{-2\epsilon}\over2(4\pi)^2}\left[{1\over\epsilon}+1\right]\;,
\\ &&
\label{lr1}
\eqa
where $\Lambda$ is the renormalization scale in the $\overline{\rm MS}$ scheme and 
$\gamma_i$ and $\delta_i$ are pure numbers,
\begin{align}
\gamma_1&={1\over3}\;,&
\quad\gamma_2&={2\over3}\;,&
\quad\gamma_3&=-{1\over2}\;,\\
\gamma_4&=2\;,&\quad\delta_1&=0\;.
\end{align}
The bare quantities are independent of the scale $\Lambda$, which implies that the
renormalized couplings satisfy simple renormalization group equations.
In the two flavor-case, it is convention to introduce the quantities $\bar{l}_i$ via
the solutions to these equations in the limit $\epsilon\rightarrow0$
\bqa
\label{lr2}
l_i^r(\Lambda)&=&{\gamma_i\over2(4\pi)^2}\left[\bar{l}_i+\log{m_{\pi,0}^2\over\Lambda^2}\right]\;.
\eqa
Up to a prefactor, $\bar{l}_i$ equals the running coupling $l_i^r$ evaluated at the scale $\Lambda=m_{\pi,0}$.
Note that since $\delta_1=0$, the coupling $h_1^r$ does not depend on $\Lambda$.
The contact term ${1\over2}h_1\langle\chi^{\dagger}\chi\rangle$ yields a constant
contribution $h_1 m_{\pi,0}^4=h_1^rm_{\pi,0}^4$ to the pressure, which will be ignored in the following.
In Sec.~\ref{cl}, we discuss the chiral limit, i.e. $m_{\pi,0}=0$. In this case,
we keep the running couplings $l_i^r(\Lambda)$.

The contribution from ${\cal L}_4^{\rm static}$ to the thermodynamic potential is
\bqa
\nonumber
\Omega_1^{\rm static}(\mu_I,\alpha)&=&
-(l_1+l_2)\mu_I^4\sin^4\alpha
-l_4m_{\pi,0}^2\mu_I^2\cos\alpha\sin^2\alpha
\\ &&
-(l_3+l_4)m_{\pi,0}^4\cos^2\alpha
\;,
\label{counter}
\eqa
The one-loop contribution to the thermodynamic potential follows directly from the inverse 
propagator. After going to Euclidean space the result is
\bqa
\nonumber
\Omega_1^{\rm loop}(\mu_I,\alpha)
&=&{1\over2}\int_P\log\left[(P^2+m_1^2)(P^2+m_2^2)+{p_0^2}m_{12}^2\right]
\\ &&
+{1\over2}\int_P\log\left[P^2+m_3^2\right]
\;.
\eqa
The pressure $\mathcal{P}$ at NLO is found by evaluating the thermodynamic potential at the LO minimum
satisfying $\cos\alpha_0={m_{\pi,0}^2\over\mu_I^2}$, as explained. This simplifies the
calculations somewhat since $m_1^2=0$.
We therefore need to evaluate 
\bqa
\nonumber
\mathcal{P}_1^{\rm loop}(\mu_I)&=&-{1\over2}\int_P\log\left[P^2(P^2+m_2^2)+{p_0^2}m_{12}^2\right]
\\ && \nonumber
-{1\over2}\int_P\log\left[P^2+m_3^2\right]\\
&=& \nonumber
-{1\over2}\int_P\log\left[P^2+m_2^2\right]
-{1\over2}\int_P\log\left[P^2+m_3^2\right]\\
&&+{1\over2}\sum_{n=1}^{\infty}{(-1)^nm_{12}^{2n}\over n}\int_P{p_0^{2n}\over P^{2n}(P^2+m_2^2)^n}\;,
\nonumber\\&&
\eqa
where we in the last line have expanded the logarithm of the first term in powers of $z=m_{12}^2/m_2^2$.
After integrating over angles in $d$ dimensions, we can write
\bqa
\nonumber
\mathcal{P}_1^{\rm loop}(\mu_I)&=&
{1\over2}I_0^{\prime}(m_2^2)+{1\over2}I_0^{\prime}(m_3^2)
+{\Gamma(2-\epsilon)\over2\Gamma({1\over2})}
\\ &&\times
\sum_{n=1}^{\infty}{\Gamma(n+{1\over2})\over\Gamma(n+2-\epsilon)}{(-1)^n m_{12}^{2n}\over n}I_n(m_2^2)\;.
\label{divp}
\eqa
We single out the two divergent terms with $n=1,2$ and resum the rest of the series~\cite{nicolis}.
In the remaining finite terms ($n\geq3$ in the sum), we can set $d=3$. 
This yields
\bqa
\nonumber
\mathcal{P}_1^{\rm loop}(\mu_I)&=&
{1\over2}I_0^{\prime}(m_2^2)+{1\over2}I_0^{\prime}(m_3^2)-{m^2_{12}\over2(d+1)}I_1(m_2^2)
\\ &&
+{3m^4_{12}\over4(d+1)(d+3)}I_2(m_2^2)
-{5m_{12}^6\over768(4\pi)^2m_2^2}
\nonumber\\&&\times
{_3F_2}
\left[\begin{array}{ccc}
1,&1,&{7\over2}\\
~&4,~5&\\
\end{array}\Bigg|-{m_{12}^2\over m_2^2}\right]\;,
\label{divp222}
\eqa
where $_3F_2$ is a hypergeometric function~\cite{grad}, 
which is given by
\bqa
{_3F_2}\left[\begin{array}{ccc}
1,&1,&{7\over2}\\
~&4,~5&\\
\end{array}\Bigg|z\right]&=&{16\over5}\left[{(3z^2-10z-8)(1-\sqrt{1-z})\over z^4}
\right.\nonumber\\&&\left.
+{z^2+4\over z^3}
-3{z^2-4z+8\over z^3}\right.\nonumber\\&&\left.\times
\log{1+\sqrt{1-z}\over2}\right]\;.
\label{divp2}
\eqa
We briefly discuss
hypergeometric functions in the Appendix.
Combining Eq.~(\ref{lop}), Eq.~(\ref{counter}) and Eq.~(\ref{divp222})  together with Eqs.~(\ref{lr1}) and~(\ref{lr2}), the pressure is
\bqa
\nonumber
\mathcal{P}_{0+1}(\mu_I)
&=&
{1\over2}f^2\mu_I^2\left[1-{m_{\pi}^2\over\mu_I^2}\right]-{1\over2}f^2{m_{\pi,0}^4\over m_{\pi}^2}\left[1-{m_{\pi}^2\over\mu_I^2}
\right]
\nonumber\\ &&
+{m_{\pi,0}^8\over6(4\pi)^2\mu^4_I}\left[\bar{l}_1+2\bar{l}_2-{3\over2}\bar{l}_3-{5\over4}
\right.\nonumber\\&&\left.
+{3\over2}\log{m_{\pi,0}^2\mu^2_I\over\mu^4_I-m_{\pi,0}^4}\right]
+{\mu^4_I\over6(4\pi)^2}\left[\bar{l}_1+2\bar{l}_2+{3\over2}
\right.\nonumber\\&&\left.
+{3\over2}\log{m_{\pi,0}^{4}\over\mu^4_I-m_{\pi,0}^4}\right]
-{5m_{\pi,0}^{12}\over12(4\pi)^2(\mu^4_I-m_{\pi,0}^4)\mu^4_I}
\nonumber\\&&\times
{_3F_2}
\left[\begin{array}{ccc}
1,&1,&{7\over2}\nonumber\\
~&4,~5&
\end{array}\Bigg|-{4m_{\pi,0}^4\over\mu_I^4-m_{\pi,0}^4}\right]
\\ && 
-{m_{\pi,0}^4\over3(4\pi)^2}\left[\bar{l}_1+2\bar{l}_2-{3\over4}\bar{l}_3+{9\over8}
\right]\;,
\label{p01final}
\eqa
where we have written the argument of the hypergeometric function as $-{m_{12}^2/m_2^2}=-4m_{\pi,0}^4/(\mu_I^4-m_{\pi,0}^4)$ and added a constant such that the pressure vanishes at $\mu_I=m_{\pi}$ (in the loop corrections, $m_{\pi}^2=m_{\pi,0}^2$
to the order we are calculating).
From the pressure, one can calculate the isospin density $n_I$ and the energy density ${\cal E}$
using the standard thermodynamic relations.

We close this section by making a comparison of our $\chi$PT result for the pressure with recent lattice data~\cite{gergy5} using physical quark masses.
The pion mass is $m_{\pi}=135~\rm{MeV}$ and the pion decay constant was slightly different in the two simulations, $f_{\pi}={130\pm3\over\sqrt{2}}$ MeV and $f_{\pi}={136\pm4\over\sqrt{2}}$ MeV, respectively. The simulations are done with two different lattice spacings, $a\approx 0.22$ fm and $a\approx 0.15$ fm. The results were not continuum extrapolated.

To make a reasonable comparison with the data, we choose $m_{\pi}=135$ MeV and $f_{\pi}={133\over\sqrt{2}}$ MeV.
At LO, we identify the parameters $m_{\pi,0}$ and $f$ with the corresponding physical values.
At NLO, we need the relations between the bare parameters $m_{\pi,0}$ and $f$ and the physical
observables at NLO,
\bqa
\label{mpi}
m_{\pi}^2&=&m_{\pi,0}^2\left[1-{m_{\pi,0}^2\over2(4\pi)^2f^2}\bar{l}_3\right]\;,\\
f_{\pi}^2&=&f^2\left[1+{2m_{\pi,0}^2\over(4\pi)^2f^2}\bar{l}_4\right]\;.
\eqa
In order to invert these relations to solve for $m_{\pi,0}$ and $f$, we
need the experimental values of $\bar{l}_3$ and $\bar{l}_4$.
We also need the numerical values for the couplings $\bar{l}_1$ and $\bar{l}_2$ 
appearing in Eq.~(\ref{p01final}).
The numerical values for the couplings $\bar{l}_i$ are taken from Ref.~\cite{colangelo}
and read $\bar{l}_1=-0.4\pm0.6$, $\bar{l}_2=4.3\pm0.1$, and $\bar{l}_3=2.9\pm2.4$, and 
$\bar{l}_4=4.4\pm0.2$.  This yields $m_{\pi,0}=136.50\pm1.2$ MeV and  $f=88.35\pm1.9$ MeV,
respectively. 
The results for the pressure is shown in Fig.~\ref{fig2bose}. The LO result is given by the black dashed line and the NLO result by the red solid curve, where we have used the central values of the parameters and couplings.
Going from LO to NLO in $\chi$PT hardly changes the result for the pressure. The agreement between the prediction between $\chi$PT and lattice simulations is very good up to 
$\mu_I/m_{\pi}\approx1.5$. 
Of course, the comparison should be taken with a grain of salt since our results are for
two flavors. A more detailed comparison with three-flavor $\chi$PT 
will be presented elsewhere~\cite{allie}.

\re{
\begin{figure}[htb]
\centering
\includegraphics[width=0.48\textwidth]{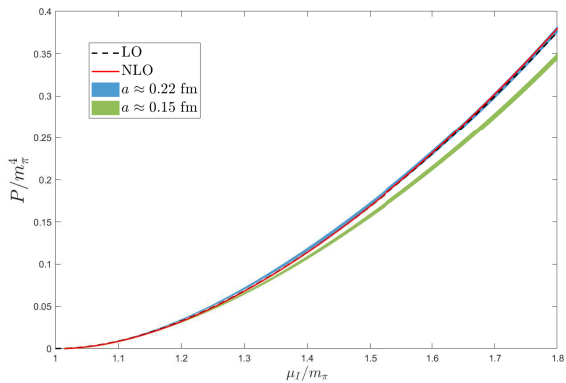}
\caption{The normalized pressure as a function of the normalized isospin chemical potential. The LO level and NLO level results are the black dashed and red solid lines, respectively. 
The blue and green bands are from two simulations
with different lattice spacings~\cite{gergy5}. See main text for details. 
}
\label{fig2bose}
\end{figure}
}

\section{Nonrelativistic limit}
In this section, we consider the nonrelativistic limit of the final expression for the pressure
Eq.~(\ref{p01final}).
In this limit, we will recover the classic results for thermodynamic quantities
for the dilute Bose gas obtained over six decades ago, which we briefly discussed in 
Sec.~\ref{dilute}. It turns out that the nonrelativistic loop corrections are hidden in the hypergeometric
function.
It will be useful to express the results in terms of the physical pion mass $m_{\pi}$
and the $s$-wave scattering length $a$, both calculated to one-loop 
and given in terms of the
bare parameters $m_{\pi,0}$ and $f$, as well as low-energy constants $\bar{l}_1$, $\bar{l}_2$, and $\bar{l}_3$~\cite{gasser1,scatgas,scat2}.
The expression we need is
\bqa
a_0^2&=&-{m_{\pi,0}^2\over4(4\pi)f^2}\left[
1-{4m_{\pi,0}^2\over3(4\pi)^2f^2}\left(\bar{l}_1+2\bar{l}_2+{3\over8}\right)\right]\;,
\label{a00}
\eqa
where the $s$-wave scattering length is $a=-a_0^2/m_{\pi}$. 
We first write the isospin chemical potential $\mu_I=m_{\pi}+\mu_{\rm NR}$, where $m_{\pi}$
is the physical pion mass as given by Eq.~(\ref{mpi}) and $\mu_{\rm NR}$ is the usual nonrelativistic
chemical potential. 
Expanding the pressure in powers of $\mu_{\rm NR}$ up to order $\mu^{5\over2}_{\rm NR}$,
using Eqs.~(\ref{mpi}) and~(\ref{a00}), we find
\bqa
{\cal P}&=&{m_{\pi}\over8\pi a}\mu^2_{\rm NR}
\left[1
-{32\over15\pi}\sqrt{4m_{\pi}\mu_{\rm NR}a^2}\right]
\;.
\eqa
The isospin density $n_I$ is found using Eq.~(\ref{densitet})
\bqa
n_I&=&{m_{\pi}\over4\pi a}\mu_{\rm NR}\left[1
-{8\over3\pi}\sqrt{4m_{\pi}\mu_{\rm NR}a^2}\right]
\;.
\eqa
Inverting this equation to find the chemical potential as a function of $n_I$, we obtain
\bqa
\mu_{\rm NR}&=&{4\pi n_I a\over m_{\pi}}\left[1+{32\over3\sqrt{\pi}}\sqrt{n_Ia^3}\right]\;.
\eqa
We can then finally calculate the nonrelativistic energy density using Eq.~(\ref{legendre}).
This yields
\bqa
\label{energynr}
{\cal E}&=&m_{\pi}n_I+{2\pi n_I^2a\over m_{\pi}}\left[
1+{128\over15\sqrt{\pi}}\sqrt{n_Ia^3}
\right]\;.
\eqa
The first term in Eq.~(\ref{energynr}) is the contribution to ${\cal E}$ associated with the
rest mass $m_{\pi}$ of the bosons.
This term is absent in Eq.~(\ref{bosongas})
since it is automatically removed by subtracting the rest mass energy in the nonrelativistic
Lagrangian Eq.~(\ref{nrlag}). The second term is the Bogoliubov mean-field term 
with loop corrections absorbed into the physical scattering length.
The last term is the LHY correction term,
where we have included these loop corrections as they are of higher orders.
Similar results were recently obtained in Ref.~\cite{newnic} for interacting scalars.

\section{Pressure to ${\cal O}(p^6)$ in the chiral limit}
\label{cl}
In the chiral limit, the calculations simplify significantly. The 
value of $\alpha$ that is a solution to the equation of motion is $\alpha={1\over2}\pi$
and the quark condensate is rotated into a pion condensate for any nonzero value
of $\mu_I$. The propagator becomes diagonal and the masses reduce to
 $m^2_{1}=0$, $m^2_{2}=\mu^2_I$, and $m^2_{3}=\mu^2_I$.
Similarly, the counterterms as well as the cubic and quartic interaction terms 
are much simpler in this limit,
\bqa
{\cal L}_2^{\rm (3)}
&=&
{\mu_I\over f}\partial_0\phi_1\left[
\phi_2^2+\phi_3^2
\right]\;,\\
\nonumber
{\cal L}_2^{\rm (4)}&=&
{1\over6f^2}\left[\phi_a\phi_b(\partial_{\mu}\phi_a)(\partial^{\mu}\phi_b)
-\phi_a\phi_a(\partial_{\mu}\phi_b)(\partial^{\mu}\phi_b)\right]
\\ &&
+{\mu_I^2\over6f^2}\phi_a\phi_a\left[\phi_2^2+\phi_3^2\right]\;,\\
\nonumber
\mathcal{L}_{4}^{\rm (2)}
  &=&-(l_{1}+l_{2}){2\mu_I^4\over f^2}\left[\phi_2^2+\phi_3^2\right]
+l_{1}{2\mu_{I}^{2}\over f^{2}}
\partial_{\mu}\phi_{a}\partial^{\mu}\phi_{a}
\\ && \nonumber
+l_{1}{4\mu_{I}^{2}\over f^{2}}
(\partial_{0}\phi_{1})^{2} 
+l_2{2\mu_{I}^{2}\over f^{2}}
(\partial_{0}\phi_{a})^{2} 
\\ &&
+l_2{2\mu_{I}^{2}\over f^{2}}
(\partial_{0}\phi_{1})^{2} 
+l_{2}{2\mu_{I}^{2}\over f^{2}}\partial_{\mu}\phi_{1}\partial^{\mu}\phi_{1}\;.
\label{l24}
\eqa
The pressure at ${\cal O}(p^4)$ follows from Eq.~(\ref{p01final}) by taking the limit $m_{\pi,0}^2\rightarrow0$ and reinstating $l_1^r$ and $l_2^r$.
This yields
\bqa
{\cal P}_{0+1}
\nonumber
&=&\frac{1}{2}f^{2}\mu_{I}^{2}+\mu_I^4
\left [{l}_{1}^r+{l}_{2}^r
+{1\over4(4\pi)^2}\left(1+2\log{\Lambda^2\over\mu_I^2}\right)\right]\;,
\\ &
\label{chiralnlo}
\eqa
where we notice that the term involving the hypergeometric function vanishes.

\begin{figure}[htb]
\includegraphics[width=0.47\textwidth]{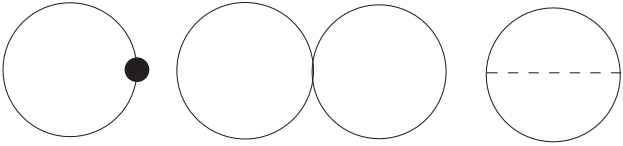}
\caption{One and
two-loop Feynman graphs contributing to the thermodynamic potential at ${\cal O}(p^6)$.
Dashed lines represent the massless Goldstone boson and solid lines represent mesons with masses $m_2=m_3=\mu_I$. The black dot represents counterterm insertions from Eq.~(\ref{l24}).}
\label{graphV2}
\end{figure}

The Feynman diagrams contributing to the thermodynamic potential at order
${\cal O}(p^6)$ are shown in Fig.~\ref{graphV2}. 
The corresponding contribution to the pressure reads
\bqa
{\cal P}_2^{\rm loops}&=&
{\mu^2_I\over6f^2}\left[3I^2_1(m^2_2)+3I^2_1(m^2_3)+2I_1(m^2_2)I_1(m^2_3)\right]
\nonumber\\&&
-{1\over6f^2}[m^2_2+m^2_3]I_1(m^2_2)I_1(m^2_3)
\nonumber\\&&
-{\mu^2_I\over f^2}\left[J(m^2_2)+J(m^2_3)\right]
\nonumber\\&&
-(l_1+l_2){2\mu^4_I\over f^2}\left[I_1(m^2_2)+I_1(m^2_3)\right]
\nonumber\\&&
+l_1{2\mu^2_I\over f^2}\left[m^2_2 I_1(m^2_2)+m^2_3 I_1(m^2_3)\right]
\nonumber\\&&
+l_2{2\mu^2_I\over f^2}\left[{m^2_2\over d+1}I_1(m^2_2)+{m^2_3\over d+1}I_1(m^2_3)\right]\;,
\label{omitted}
\eqa
where the integral $J(m^2)$ is defined in Eq.~(\ref{jdef}).
The corresponding Feynman graphs are shown in Fig.~\ref{graphV2}.
The integral $J(m^2)$ stems from the setting-sun diagram. Note also that the double-bubble graph with one massless particle and the massless bubble with
counterterm insertions
vanish in dimensional regularization since there is no mass scale in the integrals, thus the
corresponding terms are not included in 
Eq.~(\ref{omitted}).
Using $m_2=m_3=\mu_I$ and the expression for $J(m^2)$, Eq.~(\ref{jexp}), the contribution reduces to 
\bqa
\label{nnlo}
{\cal P}_2^{\rm loops}&=&{d-1\over d+1}{\mu_I^2\over f^2}I_1^2(\mu_I^2)
-l_2{4d\over d+1}{\mu_I^4\over f^2}I_1(\mu_I^2)\;.
\label{nnlop}
\eqa
Note that the $l_1$-dependence drops out.

The ${\cal O}(p^6)$ Lagrangian contains a larger number of terms, 57 for $SU(2)$ and 94 for $SU(3)$~\cite{bijn1,bijn2}. Most of them vanish in the chiral limit and for two flavors the NNLO Lagrangian reduces to
\bqa
\nonumber
{\cal L}_6&=&
C_{24}\langle(\nabla_{\mu}\Sigma^{\dagger}\nabla^{\mu}\Sigma)^3\rangle
\\ && \nonumber
+C_{25}\langle\nabla_{\rho}\Sigma^{\dagger}\nabla^{\rho}\Sigma\nabla_{\mu}\Sigma^{\dagger}\nabla_{\nu}\Sigma
\nabla^{\mu}\Sigma^{\dagger}\nabla^{\nu}\Sigma
\rangle
\\ &&
+C_{26}\langle\nabla_{\mu}\Sigma^{\dagger}\nabla_{\nu}\Sigma\nabla_{\rho}\Sigma^{\dagger}
\nabla^{\mu}\Sigma\nabla^{\nu}\Sigma^{\dagger}\nabla^{\rho}\Sigma
\rangle\;.
\label{l62}
\eqa
The contribution from the static part of ${\cal L}_6$ to the pressure is
\bqa
{\cal P}_2^{\rm static}&=&
2(C_{24}+C_{25}+C_{26})\mu_I^6\;.
\label{lstat2}
\eqa
The relation between the bare couplings $C_i$ and the renormalized
couplings $C_i^r$ is defined as
\bqa
\nonumber
C_{i}&=&{C_{i}^r\Lambda^{-4\epsilon}\over f^2}
-{\gamma_{i}^{(2)}\Lambda^{-4\epsilon}\over4(4\pi)^4f^2}
\left[{1\over\epsilon}+1\right]^2
\\ &&
+{(\gamma_{i}^{(1)}\Lambda^{-2\epsilon}(1+\epsilon)+\gamma_{i}^{(L)})\Lambda^{-2\epsilon}\over2(4\pi)^2f^2}
\left[{1\over\epsilon}+1\right]\;.
\label{cjdef}
\eqa
The coefficients needed are 
\begin{align}
\centering
\gamma_{24}^{(1)}&=-\frac{1}{(4\pi)^{2}}\frac{9}{32}\;,&
\gamma_{25}^{(1)}&=-\frac{1}{(4\pi)^{2}}\frac{67}{432}\;,\\
\gamma_{26}^{(1)}&=\frac{1}{(4\pi)^{2}}\frac{449}{864}\;,&
\gamma_{24}^{(2)}&=-{137\over72}\;,\\
\gamma_{25}^{(2)}&={5\over36}\;,&\gamma_{26}^{(2)}&={55\over72}\;,\\
\gamma^{(L)}_{24}&= -2l_1^r-{16\over3}l_2^r-{5\over4}l_6^r\;,
&\gamma^{(L)}_{25}&=2l_1^r-{1\over3}l_2^r +{1\over2}l_6^r\;,\\
\gamma^{(L)}_{26}&= {8\over3}l_2^r +{3\over4}l_6^r\;.   
\end{align}
We note that in the sum $C=C_{24}+C_{25}+C_{26}$ only $l_2^r$ remains, which is needed
to cancel the part of the divergences associated with $l_2$ in Eq.~(\ref{nnlop}). 
Adding Eq.~(\ref{nnlo}) and Eq.~(\ref{lstat2}), and renormalizing the couplings
according to Eqs.~(\ref{lr1}) and~(\ref{cjdef}), we obtain ${\cal P}_2={\cal P}_2^{\rm loops}+{\cal P}_2^{\rm static}$
\bqa
\nonumber
{\cal P}_2&=&2C_r{\mu_I^6\over f^2}-{l_2^r\mu_I^6\over2(4\pi)^2f^2}\left[1-6\log{\Lambda^2\over\mu_I^2}\right]
\\ &&
-{\mu_I^6\over24(4\pi)^4f^2}\left[1+8\log{\Lambda^2\over\mu_I^2}
-12\log^2{\Lambda^2\over\mu_I^2}
\right]\;.
\eqa
The pressure can now be written as 
\bqa
\label{pur}
{\cal P}_{0+1+2...}&=&{1\over2}f^2\mu_I^2+a_1\mu_I^4+a_2{\mu_I^6\over f^2}+\cdots\;.
\eqa
where $a_1$ can be read off Eq.~(\ref{chiralnlo}) and $a_2$ from the expression for ${\cal P}_2$.
This yields
\bqa
a_1&=&
\left [{l}_{1}^r+{l}_{2}^r+{1\over4(4\pi)^2}\left(1+2\log{\Lambda^2\over\mu_I^2}\right)\right]\;,\\
\nonumber
a_2&=&
2C_r-{l_2^r\over2(4\pi)^2}\left[1-6\log{\Lambda^2\over\mu_I^2}\right]
\\ &&
-{1\over24(4\pi)^4}\left[1+8\log{\Lambda^2\over\mu_I^2}
-12\log^2{\Lambda^2\over\mu_I^2}
\right]\;.
\label{a2}
\eqa
The coefficient $a_1$ is independent of the scale $\Lambda$, which follows from the
running of $l_1^r$ and $l_2^r$, cf. Eq.~(\ref{lr2}).
The coefficients $a_2$ does also not run, which follows from the running of $l_2^r$
and $C_r$, where the latter satisfies the renormalization group
equation in the limit $\epsilon\rightarrow0$
\bqa
\nonumber
\Lambda{dC^r\over d\Lambda}&=&{1\over6(4\pi)^4}-{3l_2^r\over(4\pi)^2}\;.
\label{dD}
\eqa
The independence of $a_i$ on $\Lambda$ guarantees that the pressure ${\cal P}$
is independent of the renormalization scale as well.
Finally, we note that the effective expansion parameter is ${\mu_I^2\over(4\pi)^2f^2}$
and the chiral limit should therefore be a good approximation 
for $m_{\pi}\ll\mu_I\ll4\pi f_{\pi}$.

\section{Discussion}
So far in this paper, we have calculated the pressure to the next-to-leading order in chiral perturbation theory. We have taken the nonrelativistic limit of our result by writing $\mu_I=m_{\pi}+\mu_{\rm NR}$ and shown that we recover Lee, Huang, and Yang's classic result for the energy
density. In the ultrarelativistic limit, we have calculated the pressure to the next-to-next-to-leading order. 

Assume now we are interested in the low-energy dynamics of the Goldstone bosons, i.e. in
momenta $p$ much smaller than the inverse coherence length in dilute Bose gases or more generally
for momenta where the linear dispersion relation is a good approximation.~\footnote{This includes the color-flavor locked (CFL) phase of dense QCD, where the momenta must be much smaller than the superconducting gap.}
In this case, Son~\cite{lowson} derived a low-energy effective theory for the superfluid phonons.
It is given in terms of the pressure ${\cal P}$ of the system as a function of the chemical potential (and possibly other quantities such as the pion mass or the quark mass) making the \sout{simply}
substitution $\mu\rightarrow\sqrt{(\partial_0\phi-\mu)^2+(\partial_i\phi)(\partial^i\phi)}$, or 
\bqa
\label{phononl}
{\cal L}_{\rm phonon}&=&{\cal P}(\sqrt{\nabla_{\mu}\phi\nabla^{\mu}\phi})\;,
\eqa
where the covariant derivative is 
$\nabla_{\mu}\phi=\partial_{\mu}\phi-\delta_{0\mu}\mu_I$.
The only approximation that was made in the derivation, is that there are as many powers of the field $\phi$ as there are derivatives. Thus the dispersion relation will be linear and the 
effective theory will break down once there are non-negligible corrections to this~\cite{lowson}. 

The conventional view of Lagrangians such as Eq.~(\ref{phononl}) is that Lorentz invariance
is explicitly broken by introducing a chemical potential as the zeroth component of a gauge field.
However, in the case of a broken charge associated with $\mu$, there is another equivalent view~\cite{nicolis0,nicolis3}.
Since $Q$ is broken, the ground state of the system is not an eigenstate of $Q$. It is not
an eigenstate of the original Hamiltonian of the system either, but it is an eigenstate of
the grand-canonical Hamiltonian ${\cal H}_Q={\cal H}-\mu Q$. Since $Q$ is broken, so are
the original time translations generated by ${\cal H}$. However, the new generator of time
translations defined by ${\cal H}_Q$ is unbroken. The ground state of the system
is now time-dependent $\sim e^{-i\mu Qt}$, which one expands about.
Using a time-dependent ground state, the system is described by the original Lorentz
invariant Lagrangian.
Clearly, the ground state breaks the original Lorentz invariance (as well as boosts), so
instead of breaking this symmetry explicitly, one may consider it being 
broken spontaneously by the 
ground state. Lorentz invariance is then realized non-linearly as is the  $U(1)$-symmetry,
where the phase transforms as $\phi\rightarrow\phi+a$.
The three broken boost generators and the broken internal $U(1)_{I_3}$ symmetry now appear, seemingly
on the same footing, in the coset construction, each multiplied by a field.
However, as pointed out in e. g.~Ref.~\cite{nicolis1}, the four fields do not represent four truly independent physical fluctuations. In fact some of them represent gauge redundancies and
are removed by a gauge choice. In the present case, there is only one physical fluctuation, as expected.
  
We will use Son's prescription to derive an effective low-energy theory for the massless mode
in dense QCD at finite isospin.
This low-energy effective theory interpolates between the nonrelativistic regime and the ultrarelativistic
regime dependent of the value of the dimensionless ratio $\mu_I/m_{\pi,0}$.
Even at the tree level, the Lagrangian contains some interesting physics.
Making the substitution 
$\mu_I\rightarrow\sqrt{(\partial_0\phi-\mu_I)^2+(\partial_i\phi)(\partial^i\phi)}$
in the LO pressure Eq.~(\ref{lop}) and 
expanding the Lagrangian in powers of derivatives and rescaling the field, we
obtain 
\bqa
{\cal L}&=&{1\over2}
(\partial_0\phi)^2-{1\over2}c_s^2(\nabla\phi)^2+c_1(\partial_0\phi)^3
+\cdots\;,
\label{expansionl2}
\eqa
where the speed of sound or phonon speed $c_s$ and the coupling $c_1$ are
\bqa
\label{cs}
c_s&=&\sqrt{\mu_I^4-m_{\pi,0}^4\over3m_{\pi,0}^4+\mu_I^4}\;,\\
c_1&=&{2m^4_{\pi,0}\mu_I\over f}{1\over(3m_{\pi,0}^4+\mu_I^4)^{3\over2}}\;.
\eqa
Note that in the chiral limit, the Lagrangian Eq.~(\ref{expansionl2}) describes a free theory.
The phonon speed interpolates between $c_s=1$ in the ultrarelativistic limit
(hence "speedy Goldstone bosons" in the title)
$m_{\pi,0}\rightarrow0$
and the nonrelativistic limit
$c_s=\sqrt{\mu_{\rm NR}\over m_{\pi,0}}$, where the latter agrees with the Bogoliubov spectrum for small $p$.
Moreover, the phonon speed $c_s$ and the coefficients $c_i$ are all subject to renormalization. The
leading corrections to the coefficients
can be found by expanding the NLO effective Lagrangian using the NLO pressure.
The speed of sound can also be found by using the thermodynamic relation
\bqa
c_s^2&=&{d{\cal P}\over d{\cal E}}\;.
\eqa
\begin{figure}[htb]
\centering
\includegraphics[width=0.48\textwidth]{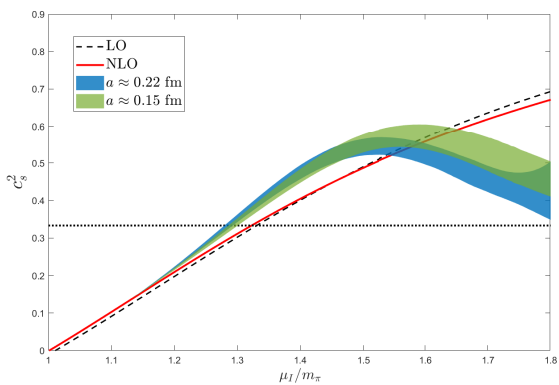}
\caption{Speed of sound squared $c_s^2$ as a function of the normalized isospin chemical potential. See. main text for details.}
\label{fig3bose}
\end{figure}
In Fig.~\ref{fig3bose}, we show the $\chi$PT results for the speed of sound squared $c_s^2$
at LO (dashed black line) and NLO (solid red line) in the low-energy expansion.
The red line corresponds to the central values of the parameters $m_{\pi,0}$, $f$, and $\bar{l}_i$ as before. 
The correction to the LO result is small in the entire region shown.
The curves approach unity in the limit $\mu_I\rightarrow\infty$.
The lattice data are of from different simulations using two different  
lattice spacings ($a\approx0.22$ fm and $a\approx0.15$ fm), respectively~\cite{gergy5}.
The bands indicate the uncertainty of the simulations.
The agreement between the predictions from $\chi$PT and the lattice results is good 
for $\mu_I$ up to approximately $1.3m_{\pi}$. At $\mu_I$ around $1.55m_{\pi}$, 
the simulations show a peak for $c_s$ and it thereafter decreases. At asymptotic values of
$\mu_I$, $c_s$ is expected to approach the conformal limit of QCD, $c_s^2={1\over3}$.
This value is shown as the horizontal dotted line in the Figure.
We note that $\chi$PT fails to reproduce the behavior found on the lattice for larger values
of the isospin chemical potential. This is expected since $\chi$PT has the wrong degrees of freedom:
In QCD at large isospin, one expects loosely bound Cooper pairs due to the attractive 
color-singlet channel in one-gluon exchange~\cite{isostep,isostep2}.~\footnote{At large $\mu_I$,
quarks are interacting weakly due to asymptotic freedom. Since the dominant interaction
is attractive, the Fermi surface is rendered unstable which leads to the formation of Cooper pairs.}

Let us finally briefly return to the effective Lagrangian and the damping of phonons.
Son's prescription
was employed in Refs.~\cite{cfl,cfl2} in the CFL phase of QCD, including scale-breaking effects due to the finite mass $m_s$ of the $s$-quark. Including the leading effects, the
phonon speed was found to be
$c_s=\sqrt{{1\over3}-{m_s^2\over9\mu_B^2}}$, where $\mu_B$ is the baryon chemical potential.
Without these effects, the phonons propagate with the
speed $c_s=1/\sqrt{3}$, which is the conformal limit of QCD.
The three-point interaction in Eq.~(\ref{expansionl2})
is of the same form as in the CFL phase, albeit with a different
coupling~\cite{cfl}.
Since the LO damping rate is calculated from the one-loop diagram arising from this term, the leading momentum dependence turns out to be the same.
In Ref.~\cite{cfl}, their damping rate was calculated in the CFL phase and it goes like $p^5$, 
for small $p$. This is exactly the same momentum dependence as the classic result by Beliav~\cite{beli}
calculated 65 years ago for the dilute Bose gas in the limit $p\rightarrow0$, which 
therefore is recovered.

\section*{Acknowledgements}
Q. Yu and H. Zhou have been supported by the Research Fund for the Doctoral Program of the Southwest University of Science and Technology under Contract No.23zx7122 and by
the Natural Science Foundation of China under Grant
No.12305091. They thank the Department of Physics at NTNU 
for their kind hospitality during their stay.
J. O. Andersen thanks Alberto Nicolis, Alessandro Podo, and Luca Santoni for useful discussions.
The authors thank B. Brandt and G. Endr\H{o}di for providing us with their recent lattice data
from Ref.~\cite{gergy5}.

\appendix

\section{Integrals}
We use dimensional regularization to regulate ultraviolet divergences. The integrals 
in Euclidean space are defined as
\bqa
\label{defintegral}
\int_P&=&
\int_{-\infty}^{\infty}{dp_0\over2\pi}\int_p\;,
\eqa
where
\bqa
\int_p&=&
\left({e^{\gamma_E}\Lambda^2\over4\pi}\right)^{\epsilon}\int{d^{d}p\over(2\pi)^{d}}\;,
\eqa
with $P=(p_0,{\bf p})$, $d=3-2\epsilon$, $p=|{\bf p}|$,
and $\Lambda$ is the renormalization scale associated with the $\overline{\rm MS}$ scheme.

The following class of one-loop integrals appears
\bqa
I_n(m^2)&=&\int_P{1\over(P^2+m^2)^{n}}\;,\\
I_{0}^{\prime}(m^2)&=&
-\int_P\log\left[P^2+m^2\right]\;,
\eqa
where $n$ is a nonnegative integer and the prime denotes differentiation with respect to $n$. 
Evaluating the integrals yields
\bqa
I_n(m^2)&=&
e^{\gamma_E\epsilon}{m^{4-2n}\over(4\pi)^2}\left({\Lambda\over m}\right)^{2\epsilon}
{\Gamma(n-2+\epsilon)\over\Gamma(n)}\;.
\eqa
$I_0^{\prime}(m^2)$ is divergent and 
$I_n(m^2)$ is divergent for $n=1,2$. Expanding them to order $\epsilon$, we find
\bqa
I_0^{\prime}(m^2)&=&
{m^4\over2(4\pi)^2}\left({\Lambda\over m}\right)^{2\epsilon}
\left[{1\over\epsilon}+{3\over2}+{\cal O}(\epsilon)\right]\;,
\\ \nonumber
I_1(m^2)&=&-{m^2\over(4\pi)^2}\left({\Lambda\over m}\right)^{2\epsilon}\bigg[{1\over\epsilon}+1
+{\pi^2+12\over12}\epsilon
\\ &&
+{\cal O}(\epsilon^2)
\bigg]
\;, 
\\ 
I_2(m^2)&=&{1\over(4\pi)^2}\left({\Lambda\over m}\right)^{2\epsilon}\left[{1\over\epsilon}
+{\cal O}(\epsilon)
\right]\;. 
\eqa
We also need to evaluate the setting-sun type integral appearing in the NNLO calculation of the
pressure in the chiral limit,
\bqa
\nonumber
J(m^2)&=&\int_{PQ}{p^2_0\over P^2(Q^2+m^2)[(P+Q)^2+m^2]}\\
&=&\int_P{p_0^2\over P^2}\Pi(P)\;, 
\label{jint}
\label{jdef}
\eqa
where we have defined the self-energy 
\bqa
\Pi(P)=\int_Q{1\over (Q^2+m^2)[(P+Q)^2+m^2]}\;. 
\eqa
Using Feynman parameters and averaging over angles, we obtain
\bqa
J(m^2)&=&
{1\over d+1}I^2_1(m^2)\;.
\label{jexp}
\eqa
In the theory of dilute Bose gases, the following integrals appear~\cite{braaten}
\bqa
I_{m,n}(M^2)&=&\int_p{p^{2m}\over p^n\left(p^2+M^2\right)^{{n\over2}}}\;. 
\label{defigam}
\eqa
They satisfy the recursion relation
\bqa
{dI_{m,n}\over dM^2}&=&-{1\over2}nI_{m+1,n+2}(M^2)\;,
\eqa
which follows directly from the definition Eq.~(\ref{defigam}).
Evaluating the integrals in dimensional regularization, we find
\bqa
\nonumber
I_{m,n}(M^2)&=&{e^{\gamma_E\epsilon}}{M^{3+2m-2n}\over(4\pi)^{{3\over2}}}\left({\Lambda\over M}\right)^{2\epsilon}
\\
&& \nonumber
\times
{\Gamma({3\over2}-{n\over2}+m-\epsilon)\Gamma(n-m-{3\over2}+\epsilon)\over\Gamma({n\over2})
\Gamma({3\over2}-\epsilon)}\;.
\\ &&
\eqa
We specifically need
\bqa
\label{i01}
I_{0,-1}(M^2)&=&{16\over15}{M^5\over(4\pi)^2}
\left[1+{\cal O}(\epsilon)\right]
\;,  \\
I_{1,1}(M^2)&=&{16M^3\over3(4\pi)^2}\left[1+{\cal O}(\epsilon)\right]\;. 
\label{i11}
\eqa
Note that both integrals are finite in $d=3$.

In the nonrelativistic limit, a generalized hypergeometric function $_3F_2$ appears.
Hypergeometric functions function of type $_pF_q$ are analytic functions of a single variable $z$
with $p+q$ parameters.
The generalized hypergeometric function has a power series representation in $z$
\bqa
\nonumber
{_pF_q}
\left[\begin{array}{c}
\alpha_1,\alpha_2,...\alpha_p\\
~\beta_1,~\beta_2,...\beta_q\\
\end{array}\Bigg| z\right]
&=&\sum_{n=0}^{\infty}{(\alpha_1)_n(\alpha_2)_n...(\alpha_p)_n\over(\beta_1)_n(\beta_2)_n...(\beta_q)_nn!}z^n\;,
\\ &&
\eqa
where $(a)_b$ is the Pochhammer symbol
\bqa
(a)_b&=&{\Gamma(a+b)\over\Gamma(a)}\;.
\eqa

\end{document}